\RequirePackage{fix-cm}
\documentclass[smallextended]{svjour3}       
\smartqed  
\usepackage{graphicx}
\usepackage{amssymb,amsmath,amsfonts,graphicx,latexsym,color,braket}
\usepackage{notes2bib}
\begin{document}

\title{Derive the Born's rule from environment-induced stochastic dynamics
of wave-functions in an open system
}


\author{Pei Wang
}


\institute{Pei Wang \at
              Department of Physics, Zhejiang Normal University, Jinhua 321004, China \\
              \email{wangpei@zjnu.cn}           
}

\date{Received: date / Accepted: date}

\maketitle

\begin{abstract}
The lack of superposition of different position states or the emergence of classicality
in macroscopic systems have been a puzzle for decades. Classicality
exists in every measuring apparatus, and is the key for understanding
what can be measured in quantum theory. Different theories
have been proposed, including decoherence, einselection
and the spontaneous wave-function collapse,
with no consensus reached up to now.
In this paper, we propose a stochastic dynamics for the wave-function in
an open system (e.g. the measuring apparatus) that interacts
with its environment. The trajectory of wave-function is random with
a well-defined probability distribution.
We show that the stochastic evolution results in the wave-function collapse
and the Born's rule for specific system-environment interactions.
While it reproduces the unitary evolution governed by the
Schr\"{o}dinger equation when the interaction vanishes.
Our results suggest that it is the way of system interacting with
environment that determines whether quantum superposition dominates
or classicality emerges.
\end{abstract}

\section{Introduction}

Quantum mechanics is exceedingly successful in explaining the experiments,
with no conflict being found up to now. But ever since its born,
the theory is debated for its inability to explain the familiar
classical world where the measuring apparatus and observers
live in. Einstein criticized that the theory does not "decide what can be observed".
Various theories have been proposed to fix this problem,
including Bohmian mechanics~\cite{Bohm},
many-worlds interpretation~\cite{Everett}, decoherence and
einslection~\cite{Zurek03}, and spontaneous wave-function collapse models~\cite{Bassi13}.

The debate stems from an axiom of quantum mechanics -
the Born's rule. It states~\cite{Nielsen} that a measurement
(a projection-valued measurement) in mathematics defines an
orthogonal basis of the Hilbert space, into which the wave-function
is decomposed: 
\begin{equation}
\ket{\phi}= \sum_s \alpha_s \ket{\phi_s}.
\end{equation}
The probability of the wave-function collapsing into $\ket{\phi_s}$
is $P_s=\left| \alpha_s\right|^2$. $\{\ket{\phi_s}\}$ is the eigenbasis of
the observable operator. But quantum mechanics
does not clarify how $\{\ket{\phi_s}\}$ is connected to the measuring apparatus.
Imagine an experiment in which we need to measure the position of
an electron. Why do we know that we are measuring the position,
but not something else like the momentum?
Whether we are measuring the position or momentum
is determined by the measuring apparatus.
Therefore, quantum mechanics cannot be self-consistent
without giving explicitly how to determine $\{\ket{\phi_s}\}$
from the description of measuring apparatus.

This problem was noticed in the early days of quantum mechanics.
In the Copenhagen interpretation~\cite{Bohr28}, there is a border between
quantum and  classical worlds. The microscopic world obeys
quantum laws. But the macroscopic world including
the measuring apparatus and observers has to
be described classically. A classical world is necessary
for determining whether the position or momentum is measured.
This is of course unsatisfying, because the border is defined ambiguously.
Many different approaches have been proposed to replace the Copenhagen interpretation.
For each approach, how to derive the Born's rule is a crucial question to answer.

A proof of the Born's rule was given by Gleason in 1957~\cite{Gleason}. But it is purely mathematical without
addressing the foundation of quantum mechanics. Everett~\cite{Everett} proposed the many-worlds interpretation,
in which both the microscopic system and macroscopic measuring apparatus
are parts of the universe with the latter being described by a pure wave function. This wave function
evolves according to the Schr\"{o}dinger equation, being totally deterministic.
To answer why the specific set of vectors
$\{\ket{\phi_s}\}$ are chosen during the measuring process,
decoherence~\cite{Zeh70} and einselection~\cite{Zurek81,Zurek82} were introduced.
Within a measurement, the microscopic system is entangled with
the measuring apparatus. At the same time,
the interaction with environment singles out a preferred basis
in the Hilbert space of the measuring apparatus, dubbed the pointer states.
The pointer states remains untouched in the interaction with environment,
while their superpositions decohere. And the environment
acquires and transmits redundant information
about the pointer states~\cite{Zurek03}.
The pointer states correspond to the classical states in our familiar world,
forming the preferred basis in the Born's rule. To answer why the probability is
$\left| \alpha_s\right|^2$, Deutsch~\cite{Deutsch} suggested an argument based on the decision theory.
His argument was debated by Barnum~\cite{Barnum} but then developed
by Saunders~\cite{Saunders} and Wallace~\cite{Wallace03,Wallace07}.
Later on, Zurek~\cite{Zurek03PRL,Zurek05,Schlosshauer} proposed a different derivation of the Born's rule based on
the environment-assisted invariance. Either Deutsch's or Zurek's arguments
remain controversial up to today~\cite{Rae,Herbut,Mandolesi18,Mandolesi19,Mandolesi20}.
The many-worlds interpretation and the derivation of the Born's rule based on it need no modification
to quantum mechanics, and are then welcome because
quantum theory has been verified by many experiments.
However, it does not really explain how the superposition
between measuring apparatus and environment is destroyed
in a single measurement~\cite{Bassi13}.
And one has to solve the Schr\"{o}dinger equation and know the
universe wave-function in the distant future
before defining the pointer states at present.
It is not clear whether such-defined pointer states exist
when the interaction between measuring apparatus
and environment is as complicated as
in real world, or how to define the pointer states
when the interaction is changing with time.

The hidden-variable theories circumvent the quantum measurement problem by
assuming the existence of pre-quantum (sub-microscopic)
degrees of freedom which are deterministic~\cite{Hooft,Genovese}. The pre-quantum state evolves classically
with the quantum mechanics being its statistical description. The Bohmian mechanics~\cite{Bohm52a,Bohm52b}
is a hidden-variable theory. According to it, the distribution of the pre-quantum states
will always obey the Born's rule once if the initial positions of the particles have a Born-rule distribution.
Recently, the timescale for dynamical relaxation to the Born's rule was analyzed
as the initial distribution deviates from it~\cite{Towler}. And new hidden-variable theories
were suggested (see Ref.~\cite{Khrennikov14} for an example). But how to observe the pre-quantum degrees of freedom
remains a problem in these theories.

For explaining the emergence of measuring apparatus' classicality,
the spontaneous collapse models modify the quantum mechanics in a way so that the
wave-function collapse is spontaneous and objective.
The Ghirardi-Rimini-Weber (GRW) model~\cite{Ghirardi86},
quantum mechanics with universal position localization~\cite{Bassi05},
and continuous spontaneous localization~\cite{Pearle89,Ghirardi90} (CSL) model
are in this family. In these models, the wave-function evolution
does not follow the deterministic Schr\"{o}dinger equation any more, but is stochastic.
For a particle of macroscopic mass,
the stochastic equation results in spontaneous localization in real space. The solution
is a Gaussian wave package~\cite{Bassi05}
with its center having a random trajectory. If the initial state
is a superposition of wave packages at different positions, to which position
the wave-function collapses into is probabilistic. And the distribution
was shown to obey the Born's rule. The spontaneous collapse
models are self-consistent and falsifiable. Due to the development of
optomechanics and matter-wave interferometry,
experiments have been carried out to put bounds on the parameters of
spontaneous collapse models~\cite{Bassi13,Arndt14,Pontin19}.
And researchers keep on raising new proposals for testing them~\cite{Yin13,Bahrami14,Nimmrichter14,Diosi14,Goldwater15}.
But there exist difficulties in generalizing these models to a relativistic theory,
which limit their application~\cite{Jones19}.

There were attempts to explain the classicality of the measuring apparatus in
the context of orthodox quantum mechanics. Kraus~\cite{Kraus81,Kraus85} proposed a quantum-mechanical model of 
a macroscopic counter that monitors the decay of an unstable particle. His model
was improved by Braun~\cite{Braun92} and Urbanowski~\cite{Urbanowski93}. In the quantum optics community, the quantum
trajectory theory~\cite{Brun02,Jacobs06} was developed to describe open quantum systems subjected to continuous measurement.
The quantum trajectory formalism, which is based on a continuous stochastic process, was
shown to be consistent with the Born's rule~\cite{Patel17}. But the theory does not really solve the fundamental
measurement problem. Instead, it shifts the quantum-classical border to the dynamics of the amplifier.

In addition to above approaches, some authors derived the Born's rule from the time
reversal symmetry~\cite{Ilyin15}, or proved that the alternatives to Born's rule violate the compositional principle of purification and
local tomography~\cite{Galley16,Galley18}. The experimental test of the Born's rule with three-path interference
was also proposed and carried out~\cite{Sorkin94,Sinha10,Quach17,Pleinert20}.

Up to now, the quantum measurement problem and the origination of the Born's rule
are not solved yet. New approach is still worth trying.
In this paper, we propose an environment-induced stochastic
dynamics of wave-functions. In our assumptions, the wave-function
collapse is objective. The evolution of wave-function in an open system is a stochastic
process but not deterministic. The Born's rule is not an axiom,
instead, it must be derived as a result of the stochastic dynamics.
At this point, we agree with the spontaneous
collapse models. On the other hand,
the wave-function collapse is due to the interaction with environment
(environment-induced), thereafter, an isolated system experiences
no wave-function collapse and follows
the same unitary evolution governed by the Schr\"{o}dinger equation. We do not need to worry about
a conflict to experiments that have been well explained by orthodox quantum mechanics for isolated systems.
At this point, we inherit the spirit of einselection.

The paper is arranged as follows. We
propose the stochastic dynamics and discuss
its properties in Sec.~\ref{sec:theory}.
In Sec.~\ref{sec:model}, we introduce the central spin model.
Sec.~\ref{sec:Born} explains how the Born's rule
arises from the stochastic dynamics. Sec.~\ref{sec:numerics}
presents the numerical results and discusses
the conditions of the Born's rule. Sec.~\ref{sec:con} summarizes
our model and results.

\section{Environment-induced stochastic dynamics of wave-functions}
\label{sec:theory}

In this section, we propose the stochastic dynamics of wave-functions in an open system.
When we say system in this paper, we mean a measuring apparatus in which
one can observe the collapse of wave-functions
and the emergence of classicality.

The system and its environment combine into an
isolated universe, whose evolution is
governed by the Schr\"{o}dinger equation. The total Hamiltonian is
\begin{equation}
\hat H = \hat H_E + \hat H_S + \hat V_{SE},
\end{equation}
where $\hat H_E$ and $\hat H_S$ are the Hamiltonians
for environmental and system's degrees of freedom, respectively,
and $\hat V_{SE}$ is the interaction between them.
We use $\mathcal{H}_E$ and $\mathcal{H}_S$ to denote the Hilbert space
for the environment and system, respectively,
and $\ket{n}$ to denote the eigenstate of $\hat H_E $ with
the energy $E_n$. Here we assume no degeneracy
and the set $\left\{\left(\ket{n},E_n\right)\right\}$ is uniquely defined,
which is a common simplifying assumption for the environment that avoids the
complications associated to continuous spectra and degeneracy.

Now suppose that at some given time $t_0$, the wave-function of system
is $\ket{\phi}$. We hope to predict the distribution of
wave-functions at arbitrarily later time $t$. For this purpose,
we need also know the wave-function of the environment at $t_0$. The environment
is not necessarily in a pure state, but can also be in a mixed state.
Without loss of generality, we suppose that the environment
is in the state $\ket{\gamma}$ with the probability $f_\gamma$.
The distribution $f_\gamma$ reflects the uncertainty in our knowledge of the environment.
The environment is in a pure state if $f_\gamma$ is a $\delta$-function.

The main assumption in this paper is that,
at the time $t\geq t_0$, the wave-function of the system is
\begin{equation}\label{eq:wave}
\ket{\phi_{n,\gamma} (t)} = \frac{1}{\sqrt{g_{n,\gamma}}}
\bra{n} e^{-i \left(t-t_0\right) \hat H} \ket{\gamma}\otimes
\ket{\phi}
\end{equation}
with the probability $P_{n,\gamma}$. Here $n$ and $\gamma$ denote
the final and initial environmental states, respectively. Notice that $\ket{\gamma}\otimes
\ket{\phi}$ is indeed the wave-function of the whole universe at $t_0$,
and $e^{-i \left(t-t_0\right) \hat H} \ket{\gamma}\otimes \ket{\phi}$ gives
the trajectory of the universe wave-function in the Hilbert space $\mathcal{H}_E \otimes
\mathcal{H}_S$. The nomalization factor reads
$g_{n,\gamma}(t)= \left\| \bra{n} e^{-i \left(t-t_0\right) \hat H} \ket{\gamma}\otimes
\ket{\phi} \right\|^2$, where $\left\| \cdot\right\|$ denotes the inner product of a vector
with itself. The probability distribution is expressed as
\begin{equation}\label{eq:P}
P_{n,\gamma}(t) = f_\gamma \ g_{n,\gamma}(t).
\end{equation}
The set of pairs $\left\{ \left( \ket{\phi_{n,\gamma} (t)}, P_{n,\gamma}(t) \right) \right\}$
defines a stochastic process in the Hilbert space $\mathcal{H}_S$.
The system's wave-function has a random trajectory.
It can take different possible values $\ket{\phi_{n,\gamma} (t)}$ at a certain time,
where $n$ and $\gamma$ are variables.
The probability $P_{n,\gamma}(t)$ must be
normalized at arbitrary $t\geq t_0$. It is easy to verify
$\sum_{n,\gamma} P_{n,\gamma}=1$.

A physically well-defined evolution should be continuous with time. The above
stochastic process is indeed continuous. Especially, at $t=t_0$,
Eqs.~(\ref{eq:wave}-\ref{eq:P}) tell us $\ket{\phi_{n,\gamma} (t)} \equiv \ket{\phi}$
(expect for an unphysical global phase) for arbitrary $n$ and $\gamma$,
thereafter, the system is in the state $\ket{\phi}$ with $100\%$ probability.
This is consistent with our initial condition.

For an isolated system ($\hat V_{SE}=0$), Eqs.~(\ref{eq:wave}-\ref{eq:P}) tell us that
the wave-function is $e^{-i(t-t_0)\hat H_S} \ket{\phi}$
with $100\%$ probability except for an unphysical global phase.
This is exactly the solution of the Schr\"{o}dinger
equation. For an open system ($\hat V_{SE}\neq 0$), we also find a
correspondence between quantum mechanics and our model.
Eqs.~(\ref{eq:wave}-\ref{eq:P}) give the distribution of the wave-functions.
On the other hand, quantum mechanics predicts
the reduced density matrix at a given time
by tracing out the environmental degrees of freedom, which is
$\hat \rho_S(t) = \text{Tr}_E \hat \rho(t)$ with the universe
density matrix expressed as $ \hat \rho(t)= \sum_{\gamma}f_\gamma \ket{\Psi_\gamma(t)}
\bra{\Psi_\gamma(t)} $ and the universe wave function as
$\ket{\Psi_\gamma(t)} = e^{-i \left(t-t_0\right) \hat H} \ket{\gamma}\otimes
\ket{\phi}$. From Eqs.~\eqref{eq:wave} and~\eqref{eq:P}, it is easy to prove
\begin{equation}\label{eq:dm}
\hat \rho_S(t) = \sum_{n,\gamma} P_{n,\gamma}(t) 
\ket{\phi_{n,\gamma} (t)} \bra{\phi_{n,\gamma} (t)}.
\end{equation}
Eq.~\eqref{eq:dm} is just the von Neumann's definition of density matrix.
Therefore, our equations predict the same reduced density matrix as quantum mechanics,
but contain more information,
because the decomposition of $\hat \rho_S(t)$ into an ensemble of wave functions
is not unique (the basis can be chosen arbitrarily).
This non-uniqueness reflects the disability of quantum mechanics
in explaining the emergence of classical states.

We like to mention that $\ket{\phi_{n,\gamma} (t)}$ at different $\left(n,\gamma\right)$
are not orthogonal to each other. Therefore, they do not form a basis
of $\mathcal{H}_S$, and must be distinguished from the pointer states
in einselection.

It is worth emphasizing that our assumption relies on a non-symmetric prescription for the evolution of environment
and system. The dynamical law~\eqref{eq:wave} has to distinguish between the environment
on one hand and the system on the other. According to our assumption, the environment
contains the observer. For a specific observer, he can always tell which party
is the system and which is the environment. The trajectory of the system's state is interesting to the observer,
but that of the environment is not.

\section{Central spin model}
\label{sec:model}

For a generic Hamiltonian, the wave function and probability defined by Eqs.~\eqref{eq:wave} and~\eqref{eq:P} are difficult to
calculate. To obtain a paradigm of the stochastic process defined above, we consider the system to
be as simple as a spin-$1/2$ particle. And we can neglect the system's Hamiltonian, if we only study
the collapse of its wave function. The environment has much more
degrees of freedom than the system. To keep the model solvable, we treat the environment as
a collection of $N$ spins without interaction between each other. The environmental spins and
the system spin have an interaction, which serves as the reason for the wave-function collapse.

Such a model is often called the central spin model. The model studied next
is similar to that studied in Ref.~\cite{Zwolak16}. The environmental Hamiltonian is
\begin{equation}
\hat H_E = \mu \sum_{j=1}^N \hat \sigma_j^z,
\end{equation}
where $\hat \sigma_j^z$ is the Pauli matrix of the $j$-th spin.
It is equivalent to say that all the environmental spins
are in $z$-direction for an eigenstate.
No generality is lost, because we can always define
the direction of each environmental spin in its own inner space
as the "$z$-direction".
The eigenstate of $\hat H_E$ is written as $\ket{n}=\ket{s_1, \cdots s_N}$ with
$s_j= \pm 1$ denoting the spin-up and spin-down states, respectively.
The interaction between system and environment is
\begin{equation}
\hat V_{SE} = \sum_{j=1}^N   \hat \sigma_S^z \left(h_j \hat \sigma_j^x + \nu \hat \sigma_j^z \right),
\end{equation}
where $\hat \sigma_S^z$ is the Pauli matrix of the central spin.
The coupling is in the longitudinal direction
together with a component $h_j$ in the transversal direction.

We set $t_0=0$ as the initial time. And the environment
is in thermal equilibrium, i.e., the environment is in the state $\ket{n}$ with
probability $f_n = e^{-\beta E_n}/Z$ with $Z= \sum_n e^{-\beta E_n}$ the
partition function and $\beta $ the inverse of temperature.
The dynamics that we are interested in
is indeed independent of $f_n$ and $\beta$.
The initial wave-function of the system is generally expressed as
\begin{equation}
\ket{\phi} = \alpha_\uparrow \ket{\uparrow} + \alpha_\downarrow \ket{\downarrow},
\end{equation}
where $\ket{\uparrow}$ and $\ket{\downarrow}$ are the eigenstates
of $\hat \sigma_S^z$ and $\left|  \alpha_\uparrow\right|^2
+\left|\alpha_\downarrow\right|^2=1$ is the normalization condition.
Here $ \ket{\uparrow}$ and $\ket{\downarrow}$ are the two classical
states. According to the Born's rule,
the wave-function of the system should collapse into them
with the probabilities $\left|  \alpha_\uparrow\right|^2$ and
$\left|\alpha_\downarrow\right|^2$, respectively.

Following Eqs.~(\ref{eq:wave}-\ref{eq:P}),
we obtain the distribution of wave-functions at arbitrary $t>t_0$. Using
$\left(n',n\right)=\left( \left(s'_1,s_1\right), \cdots, \left(s'_N,s_N\right)\right)$
to denote a pair of initial and final environmental states, we find the system's wave function
to be
\begin{equation}\label{eq:spinwave}
\ket{\phi_{n',n}(t)} = \frac{\displaystyle \alpha_\uparrow G^\uparrow_{n',n} }{\sqrt{g_{n',n}(t)}}
\ket{\uparrow} + \frac{\displaystyle \alpha_\downarrow G^\downarrow_{n',n} }{\sqrt{g_{n',n}(t)}}
\ket{\downarrow}
\end{equation}
with the probability $P_{n',n} (t)=f_n g_{n',n}(t)$. The normalization factor is
\begin{equation}
g_{n',n} = \left| \alpha_\uparrow \right|^2 \left| G^\uparrow_{n',n}\right|^2+
 \left| \alpha_\downarrow \right|^2 \left| G^\downarrow_{n',n}\right|^2,
\end{equation}
where $G^{\uparrow/\downarrow}_{n',n} = \displaystyle\prod_{j=1}^N 
G^{\uparrow/\downarrow}_j$ are the spin-up and spin-down
significances, respectively. And $G^{\uparrow/\downarrow}_j = \bra{s'_j} e^{-it \left(
\left(\mu\pm\nu\right)\hat \sigma^z_j \pm h_j \hat \sigma^x_j \right) }\ket{s_j} $
is the significance contributed by the $j$-th environmental degree of freedom.
Note that $\left| G^{\uparrow/\downarrow}_j \right|^2$ depends only upon
$s_js'_j$, but is independent of the initial environmental state.

\subsection{Solutions without wave-function collapse}

Let us first consider special choices of parameters to obtain
an analytical solution.
If $h_j \equiv 0$ and the environment is at zero temperature, the wave-function is
$\ket{\phi(t)}= \alpha_\uparrow e^{-it E_{gs}} \ket{\uparrow}+
\alpha_\downarrow e^{itE_{gs}} \ket{\downarrow}$ with $100\%$ probability,
where $E_{gs}$ is the ground-state energy of the environment.

Another choice is $\nu=0$ and $h_j\equiv h$ being a constant.
Because the environmental degrees of freedom
(the universe) are much more than the system's,
it is reasonable to take the large-environment limit ($N\to \infty$).
In this limit and for almost all the time,
the system's wave-function is $\alpha_\uparrow\ket{\uparrow}+
\alpha_\downarrow\ket{\downarrow}$ or
$\alpha_\uparrow\ket{\uparrow}- \alpha_\downarrow\ket{\downarrow}$
with equal probabilities. Only at the specific times
$t=m\pi/\sqrt{\mu^2+h^2}$ with $m$ an arbitrary integer,
the wave-function recovers to its initial value with $100 \%$ probability.
In both choices, there is no wave-function collapse, and
the superposition principle of quantum mechanics is preserved.

\section{The Born's rule}
\label{sec:Born}

In this section, we show how the classicality and the Born's rule emerges
from Eqs.~(\ref{eq:wave}) and~(\ref{eq:P}) in the central spin model.
Recall that the spin-up significance is $\left| G^\uparrow_{n',n} \right|^2
=\prod_{j=1}^N \left| G^\uparrow_j\right|^2$
and the spin-down significance is $\left| G^\downarrow_{n',n} \right|^2
=\prod_{j=1}^N \left| G^\downarrow_j\right|^2$.
We choose $\mu+\nu=1$ as the unit of energy and its inverse the unit of time.
The significance on the $j$-th environmental degree of freedom is
\begin{equation}\label{eq:sigspin}
\left| G^S_j\right|^2= \bigg\{ 
\begin{array}{cc} \cos^2\left(\omega^S_j t\right)
+\sin^2\left(\omega^S_jt\right)  r^S_j, & d_j=1 \\[5mm]
\sin^2\left(\omega^S_j t \right) \left(1-r^S_j\right), & d_j =-1.
\end{array}
\end{equation}
Here $S=\uparrow,\downarrow$ denotes the spin-up
and spin-down states of the system, respectively. $d_j=s_js'_j=\pm 1$
represents flip or no-flip of the $j$-th environmental spin, respectively.
$\omega^\uparrow_j= \sqrt{1+h^2_j}$ is the frequency and $r^\uparrow_j=
1/\left(1+h_j^2\right)$ is the ratio for spin-up state, while
$\omega^\downarrow_j= \sqrt{\left(\mu-\nu\right)^2+h^2_j}$
and $r^\downarrow_j= \left(\mu-\nu\right)^2/\left(\left(\mu-\nu\right)^2+h_j^2\right)$
are for spin-down state.

\begin{figure}[tbp]
\vspace{0.5cm}
\includegraphics[width=0.9\linewidth]{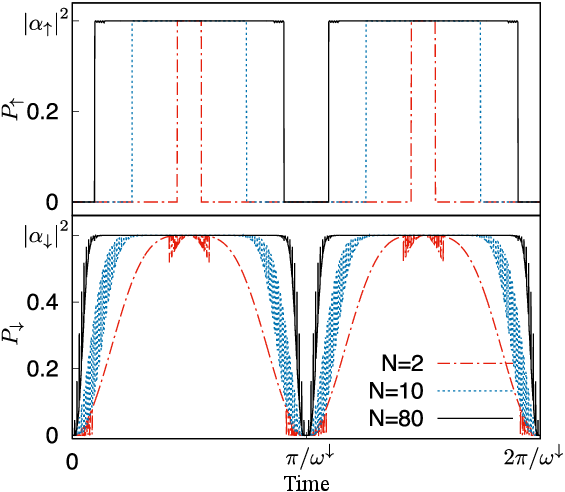}
\caption{(Color online) The evolution of $P_\uparrow$ and
$P_\downarrow$ for different $N$. We choose $\mu-\nu=0$, $h_j=0.01$ being a constant,
and $\left| \alpha_\uparrow\right|^2=0.4$. The error is $\epsilon=10^{-3}$.}\label{fig:N}
\end{figure}
We notice $\left| G_j^{\uparrow} \right|^2_{d_j=1}+
\left| G_j^{\uparrow} \right|^2_{d_j=-1} \equiv 1$.
With $j$ varying from $1$ to $N$, there are $N$ pairs of positive numbers with each
pair summing up to unity. The spin-up significance is obtained by
choosing one number from each pair according to whether the involved degree
of freedom is flipped or not, and then multiplying all the $N$ numbers.
The spin-down significance is obtained in the same way.
If $N$ is large, depending on how unity is
decomposed into two positive numbers for each $j$,
the ratio of spin-up to spin-down significance
can be close to zero or very large (think about the fact
$\left( 0.9\times 0.1\right)^{N/2} / \left( 0.5\times 0.5 \right)^{N/2} \approx 0 $).
If this happens, Eq.~\eqref{eq:spinwave} tells us that $\ket{\phi_{n',n}(t)}$
is either $\ket{\uparrow}$ or $\ket{\downarrow}$,
but cannot be a superposition of $\ket{\uparrow}$ and $\ket{\downarrow}$.
The wave-function collapse is then realized.

Let us consider the case $\left| \mu-\nu\right|/\left| h_j\right| \ll 1$
and $\left| h_j\right| \ll 1$. For this choice of parameters, we have
$r^\downarrow_j \to 0 $ and $r^\uparrow_j \to 1$.
Note that $\left| G^S_j\right|^2$ changes periodically with time
with the lower bound of $\left| G^S_j\right|^2_{d_j=1}$ being $r_j^S$
and the upper bound of $\left| G^S_j\right|^2_{d_j=-1}$ being $1-r_j^S$.
For $r^\uparrow_j = 1$, $\left| G^\uparrow_j\right|^2$
is always $1$ for non-flipped environmental degree of freedom
but $0$ for flipped one. And their product
(spin-up significance) is always $0$ except for the case
$n'=n$ (no one flipped) in which the significance is $1$.
For the more general case that $r^\uparrow_j $ is
close but not equal to $1$, we obtain similar results
once if $N$ is large. Now $\left| G^\uparrow_j\right|^2$ at ${d_j=\pm 1}$
both display small oscillations, but their product is less than $\left(1/2\right)^2$
for most of the time.
If $N$ is large, for almost all pairs of $\left(n',n\right)$,
there are approximately half of environmental spins flipped and half non-flipped.
The product of $\left| G^\uparrow_j\right|^2$ then satisfies
$\left| G^\uparrow_{n',n} \right|^2 \ll \left(1/2\right)^{N}$
except for $n'=n$ at which we have $\left| G^\uparrow_{n',n} \right|^2 
\gg \left(1/2\right)^{N}$.
On the other hand, since $r^\downarrow_j$ is close to $0$,
$\left| G^\downarrow_j\right|^2$ becomes $\cos^2\left(\omega^\downarrow_j t\right)$
for non-flipped $j$-th spin but $\sin^2\left(\omega^\downarrow_j t\right)$ for flipped one.
$\left| G^\downarrow_j\right|^2$ has a strong oscillation, being neither $0$ nor $1$
for most of the time but something between.
Except for the specific times (integer times of $\pi/ \omega^\downarrow_j $),
The spin-down significance at different $n'$ is similar to each other,
being a positive number $\sim 1/2^N$ (note that $n'$ can take $2^N$ different values).
The ratio of spin-up to spin-down significance
($\left| G^\uparrow_{n',n} \right|^2 / \left| G^\downarrow_{n',n} \right|^2 $)
is approximately zero for $n'\neq n$ but infinite for $n'=n$.
We like to emphasize that $\sum_{n'} \left| G^S_{n',n} \right|^2 \equiv 1$
for $S=\uparrow$ or $\downarrow$.
Therefore, the ratio cannot be always $0$ or $\infty$.
Both cases must happen for some values of $n'$. If the ratio is
$0$ or $\infty$, Eq.~\eqref{eq:wave} tells us that the corresponding wave function
must be $\ket{\downarrow}$ or $\ket{\uparrow}$, respectively.
And the sum of $g_{n',n}$ for $n'\neq n$ must be $\left|\alpha_\downarrow\right|^2$,
and $g_{n,n}$ is $\left|\alpha_\uparrow\right|^2$. Eq.~\eqref{eq:P} then
tells us that the wave-function collapses to $\ket{\uparrow}$
with the probability $\left|\alpha_\uparrow\right|^2$,
but to $\ket{\downarrow}$ with the probability $\left|\alpha_\downarrow\right|^2$.
This is the Born's rule.

\section{Numerical results}
\label{sec:numerics}

\begin{figure}[tbp]
\vspace{0.5cm}
\includegraphics[width=0.9\linewidth]{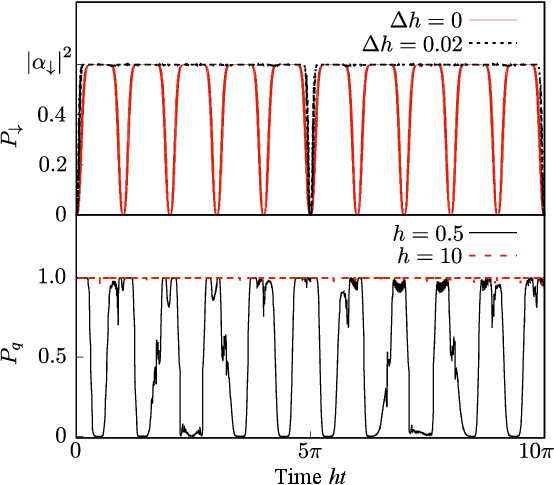}
\caption{(Color online) The evolution of $P_\downarrow$ and $P_q$.
The parameters are chosen as same as those in Fig.~\ref{fig:N},
except that $h_j$ is uniformly distributed in the range $[h,h+\Delta h)$,
expressed as $h_j= h+\left(j-1\right)\Delta h/N$. For the top panel,
we choose $h=0.01$. And for the bottom panel,
we fix $\Delta h=0.02$.}\label{fig:h}
\end{figure}
The numerical results verify above analysis and
show that the Born's rule is stable against small perturbations.
We notice that, in the stochastic dynamics,
the wave-function does not evolve exactly into the classical states
($\ket{\uparrow}$ or $\ket{\downarrow}$) for a finite $N$,
but can be infinitely close to them in the limit $N\to \infty$.
The Born's rule is only an approximation in the case of large $N$.
For numerical calculation, we define the projection
\begin{equation}
u = \left| \braket{\uparrow | \phi_{n',n}(t)} \right|^2.
\end{equation}
The projection has a distribution at a given time, denoted by $P(u)$
with $u \in [0,1]$. That $u$ is close to zero (one)
means that the wave-function is in the classical
state $\ket{\downarrow}$ ($\ket{\uparrow}$).
And we define a small positive number $\epsilon$ (dubbed the error),
which denotes how close to the classical states the wave-function is
for being considered as classical. In this spirit,
$P_\downarrow=P(0\leq u \leq \epsilon)$ is
the probability of the system being in the state $\ket{\downarrow}$,
$P_\uparrow=P(1-\epsilon \leq u \leq 1)$ is the probability
of the system being in the state $\ket{\uparrow}$.
And $P_q=1-P_\uparrow-P_\downarrow$
is the probability that the system is not in the classical states but
has to be treated as the superposition of them, i.e., the
classicality fails to emerge and the quantum superposition dominates.

Fig.~\ref{fig:N} shows the evolution of $P_{\uparrow}$
and $P_{\downarrow}$ at different $N$.
For a large $N$ ($N=80$), we see that the wave-function quickly collapses into
the classical states with the probabilities obeying the Born's rule
($P_{\uparrow/\downarrow}=\left| \alpha_{\uparrow/\downarrow}\right|^2$).
The collapsing time decreases with $N$ increasing.
The wave-function collapse
and the Born's rule have already been seen for $N=10$,
but they are absent for a smaller $N$ ($N=2$).
An interesting feature of Fig.~\ref{fig:N} is the periodic resurrection of
quantum superposition and suppression of collapse
when the time is an integer times of $\pi/\omega^\downarrow$.
This is not surprising. At these specific times, the spin-up
and spin-down significances become similar to each other because
$\sin\left(\omega^\downarrow t\right)=0$, thereafter,
the superposition between classical states resurrects.

\begin{figure}[tbp]
\vspace{0.5cm}
\includegraphics[width=0.9\linewidth]{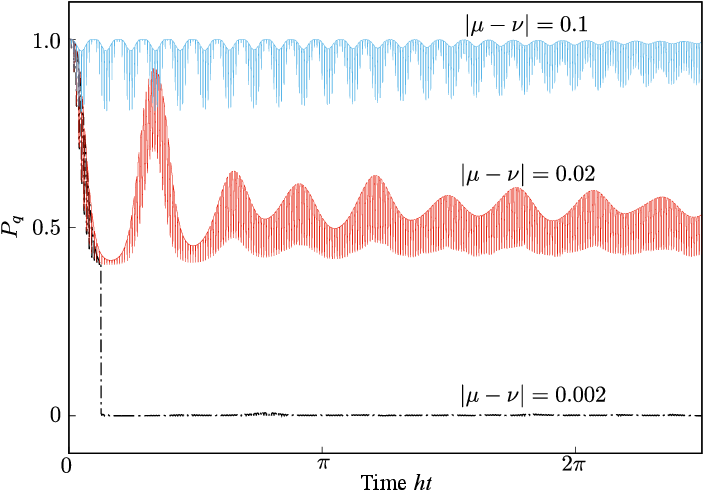}
\caption{(Color online) The probability $P_q$ at different $\left|\mu-\nu\right|$.
Here we choose $h=0.01$ and $\Delta h=0.02$.}\label{fig:munu}
\end{figure}
The resurrection of quantum superposition can be suppressed by
a dispersed $h_j$. The resurrection of superposition needs
$\left| G^\downarrow_j\right|^2_{d_j=1}=1$ and
$\left| G^\downarrow_j\right|^2_{d_j=-1}=0$.
From Eq.~\eqref{eq:sigspin}, we see that this is satisfied only
if $\omega_j^\downarrow t$ is an integer times of $\pi$. But
the frequency $\omega_j$ depends on $h_j$. A dispersed $h_j$ results
in a dispersed frequency, so that this condition is hard to be
simultaneously satisfied for different $j$, and then the resurrection
is avoided. Fig.~\ref{fig:h} (top panel)
compares $P_\downarrow(t)$ between a constant $h_j$ and a dispersed $h_j$.
We see that a dispersed $h_j$ does suppress the
resurrection by increasing its period by five times. The resurrection is still present
since there are only finite number of frequencies ($N=10$). One expects
that the resurrection will be further suppressed with $N$ increasing.

The bottom panel of Fig.~\ref{fig:h} shows the probability of
quantum superposition for different $h$. We see a strong coupling
between system and environment in the vertical direction kills
the classicality. The emergence of classicality ($P_q\approx 0$)
can be seen sometimes at $h=0.5$, but is totally lost at $h=10$.

We have argued that one condition of wave-function collapse is
$\left|\mu-\nu\right|/h \ll 1$. Fig.~\ref{fig:munu} compares
$P_q$ for different $\left|\mu-\nu\right|$, while $h$ is fixed to $0.01$.
The wave-function collapse and the Born's rule are stable
against a small $\left|\mu-\nu\right|$.
At $\left|\mu-\nu\right|=0.002$, the wave-function collapse
and the Born's rule are still clear ($P_q=0$).
But with $\left|\mu-\nu\right|$ increasing,
the classicality gradually vanishes, and the quantum effect resurrects.
When $\left|\mu-\nu\right|=0.1$ is ten times of $h$, the probability
of wave-function collapsing completely vanishes.

\section{Conclusions}
\label{sec:con}

In summary, our model of wave-function stochastic evolution
succeeds in explaining the emergence of classicality and the Born's rule
for an open system. Our main assumptions are
Eqs.~\eqref{eq:wave} and~\eqref{eq:P}, which give
the trajectories of the wave-function and their corresponding probability.
The randomness in the wave-function evolution
is environment-induced, depending on the interaction between
system and environment. And it vanishes with the interaction.
For an isolated system, Eqs.~\eqref{eq:wave} and~\eqref{eq:P} coincide
exactly with the Schr\"{o}dinger equation.

We study the stochastic dynamics in the central spin model,
showing how it results in the wave-function collapse and
the Born's rule for the system-environment interaction in a specific range.
The Born's rule is then not a priori hypothesis any more,
but an approximation in the large-environment limit.
The condition for the emergence of classicality is $\left| \mu-\nu\right|/\left| h_j\right| \ll 1$,
$\left| h_j\right| \ll 1$ and a dispersed $h_j$. If these conditions are not
satisfied, e.g., in the case $h_j=0$ or $\nu=0$,
the superposition between classical states is preserved
in the evolution, and classicality does not emerge. Therefore,
the quantum superposition is not limited in the microscopic
world, and classicality is not always present in the macroscopic
world. Whether the system keeps the quantum superposition
or shows classicality is determined by how the system
interacts with its environment. 
This is the spirit of einselection or quantum Darwinism.
But in our theory, the loss of superposition is objective, with
the process described explicitly by Eqs.~\eqref{eq:wave} and~\eqref{eq:P}.

Eqs.~\eqref{eq:wave} and~\eqref{eq:P} give the random trajectory of the wave-function
for an open system. While quantum mechanics predicts the reduced density
for an open system. There is a one-to-many map between
a reduced density matrix and the random distributions of wave-functions.
According to Eq.~\eqref{eq:dm}, the random trajectory
predicts exactly the same reduced density matrix as quantum mechanics.
But quantum mechanics by itself cannot predict the random
trajectory in an open system.
This explains why we can derive the Born's rule but it has to be an axiom
in quantum mechanics.

It is worth emphasizing the difference between our approach and previous ones,
especially the ones based on many-world interpretation or spontaneous collapse model.
In the many-world interpretation, the evolution of the physical state of the universe is
deterministic. The probability (the Born's rule) has to be introduced into the theory by
Zurek's envariance or Deutsch's decision theory.
But in our theory, the evolution of the system's wave function is defined to be stochastic
with the probability explicitly given by Eq.~\eqref{eq:P}. At this point, our theory is similar to the spontaneous collapse
models which also assume stochastic evolution. But the spontaneous collapse models are expressed in terms of
stochastic differential equations. The treatment of these equations and their generalization
to relativistic model are difficult. While in our theory, the stochastic process is defined according to
the Schr\"{o}dinger equation~\eqref{eq:wave} of orthodox quantum mechanics with which
the physicists are familiar. The price we have to pay is that the Born's rule in our theory becomes an
approximation which stands only for some specific system-environment interaction.

Therefore, we propose a model of the wave-function objectively
collapsing. It is falsifiable, because one
can use it to judge which type of system-environment
interaction can induce collapse and which cannot.
Our model might be helpful in explaining the recent experiments
on the wave-function collapse in mesoscopic systems.

\begin{acknowledgements}
This work is supported by NSFC under Grant
Nos. 11774315 and 11835011, and the Junior Associates program
of the Abdus Salam International Center for Theoretical Physics.
\end{acknowledgements}

%
%



\end{document}